\documentclass[runningheads]{llncs}
\usepackage{graphicx}
\usepackage{comment}

\usepackage{booktabs}
\usepackage{adjustbox}
\usepackage{array}
\usepackage[usenames,dvipsnames,table]{xcolor}
\usepackage{hyperref}
\hypersetup{
	colorlinks = true,
	citecolor = MidnightBlue,
	linkcolor = Maroon,
	urlcolor = Maroon
}

\definecolor{pa1}{HTML}{DE5B6D}
\definecolor{pa2}{HTML}{138086}
\definecolor{pa3}{HTML}{CBEB4E}
\definecolor{pa4}{HTML}{CD7672}
\definecolor{pa5}{HTML}{EEB462}

\usepackage{setspace}

\usepackage[ruled]{algorithm2e}
\makeatletter
\renewcommand{\@algocf@capt@plain}{above}
\makeatother
\usepackage{algorithmic}

\usepackage{wrapfig}

\usepackage{geometry}
\usepackage{tikz}
\usepackage{pgfplots}
\usetikzlibrary{pgfplots.dateplot}
\usepackage{pgfplotstable}
\usepackage{filecontents}
\usepgfplotslibrary{fillbetween}
\usetikzlibrary{patterns}
\usetikzlibrary{backgrounds}
\usetikzlibrary{arrows,automata}

\usetikzlibrary{external}
\graphicspath{ {files/} }

\begin{document}
	%
	
	\title{Moving Target Defense for Robust Monitoring of Electric Grid Transformers in\\Adversarial Environments}
	
	
	%
	\titlerunning{MTD for Robust Monitoring in Adversarial Environments}
	%
	\author{Sailik Sengupta,
		Kaustav Basu, \\
		Arunabha Sen~and~Subbarao Kambhampati
	}
	\authorrunning{S. Sengupta {\em et al.}}
	%
	\institute{
		Computing, Informatics, and Decision Systems Engineering\\
		Arizona State University\\
		\email{\{sailiks,kaustav.basu,asen,rao\}@asu.edu}}
	\maketitle              
	\begin{abstract}
		Electric power grid components, such as high voltage transformers (HVTs), generating stations, substations, etc. are expensive to maintain and, in the event of failure, replace. Thus, regularly monitoring the behavior of such components is of utmost importance. Furthermore, the recent increase in the number of cyberattacks on such systems demands that such monitoring strategies should be robust. In this paper, we draw inspiration from work in Moving Target Defense (MTD) and consider a dynamic monitoring strategy that makes it difficult for an attacker to prevent unique identification of behavioral signals that indicate the status of HVTs. We first formulate the problem of finding a differentially immune configuration set for an MTD in the context of power grids and then propose algorithms to compute it. To find the optimal movement strategy, we model the MTD as a two-player game and consider the Stackelberg strategy. With the help of IEEE test cases, we show the efficacy and scalability of our proposed approaches.\footnote{Accepted to the Conference on Decision and Game Theory for Security (GameSec), 2020.}
		
	\end{abstract}
	
	\section{Introduction}
	
	The electric power grid forms the backbone of all the other critical infrastructures (communication, transportation, water distribution, etc) of a country, and thus, necessitates the presence of adequate monitoring strategies to quickly detect any anomalous behavior(s) that may have manifested in the system. It is of utmost importance to not only detect such anomalous behavior but also to take appropriate actions quickly to prevent the failures of power grid components which in turn, may lead to a large scale blackout \cite{blackout}. Components such as High Voltage Transformers (HVTs), generating stations, substations, etc. are essential to the power grid and thus, their operational behaviors are monitored at all times with the help of Phasor Measurement Units (PMUs are devices, which are utilized as sensors, for monitoring the power grid). The problem of placing these sensors has been studied by multiple researchers over the past decade \cite{salehi2012laboratory,pal2016pmu}. Recently, in \cite{basu2018health,padhee2020identifying}, the authors proposed a sensor placement approach that can \textit{uniquely identify the source of the anomaly by utilizing the sensor readings generated by PMUs}.
	With the continuous discovery of real-world attacks such as Stuxnet \cite{karnouskos2011stuxnet}, Dragonfly \cite{team2017dragonfly} and a wide range of cyberattacks-- jamming, Denial of Service, packet dropping, false-data injection and compromise of data integrity \cite{n2020modelagnostic,niu2019framework}-- robustness of existing sensor placement mechanisms becomes critical. Thus, in this work, we leverage the ideas of Moving Target Defense (MTD) in cybersecurity \cite{jajodia2011moving,sengupta2017game} and the Minimum Discriminating Code Set (MDCS) based PMU placement \cite{basu2019sensor,basu2018health} to build a defense-in-depth solution.
	
	
	
	We continuously move the detection surface to make it challenging for an adversary to impede the unique identification of failure signals of HVTs. While PMUs are difficult to move, as opposed to the movement of physical resources in security games \cite{paruchuri2008playing}, once placed, they can be efficiently activated and deactivated, similar to the dynamic movement in intrusion detection systems \cite{sengupta2018moving}. While one may choose to activate all the PMUs placed upfront, the cost of maintaining them can become an impediment. Hence, the periodic use of a smaller subset (that still ensures unique identification) of the sensors placed upfront can be considered.
	Further, work in MTD has relied solely on heuristic guidance when constructing the configuration set that can result in all defenses being vulnerable to one attack, i.e. it is {\em not differentially immune} \cite{sengupta2019mtdeep}. In this paper, we propose methods that ensure the MTD configuration set is differentially immune.
	
	First, we define a novel variant of the MDCS problem, called the $K-$differentially Immune MDCS (hereafter $K$-$\delta$MDCS). We find $K$ MDCSs of a graph, in which all $K$ solutions can uniquely identify failing HVTs, with the added constraint that no two MDCSs share a common vertex; thus resulting in a differentially immune configuration set for the MTD. Given that the original MDCS problem is NP-Complete, we show that $K$-$\delta$MDCS is also NP-Complete and provide an optimal Quadratically Constrained Integer Linear Programming (QC-ILP) approach to find the $K_{\max}$-MDCS of a graph. While our approach proves scalable for large power networks (MATPOWER IEEE test cases), we also propose a greedy approach that is computationally faster but trades-off on finding the largest $K$ value. Second, we model the interaction between the power utility company (hereafter, the defender) and the adversary, as a normal-form game. The notion of Strong Stackelberg equilibrium used in this game-theoretic formulation, popular in existing literature \cite{sinha2015physical,sengupta2017game}, assumes a strong-threat model and aids in finding a good sensor activation strategy for the defender. Finally, we show the efficacy of our strategy and the scalability of our proposed approach on several IEEE power test cases of varying sizes.

	\section{Preliminaries}

	
	\begin{figure}
		\centering
		\includegraphics[width = 0.50\textwidth]{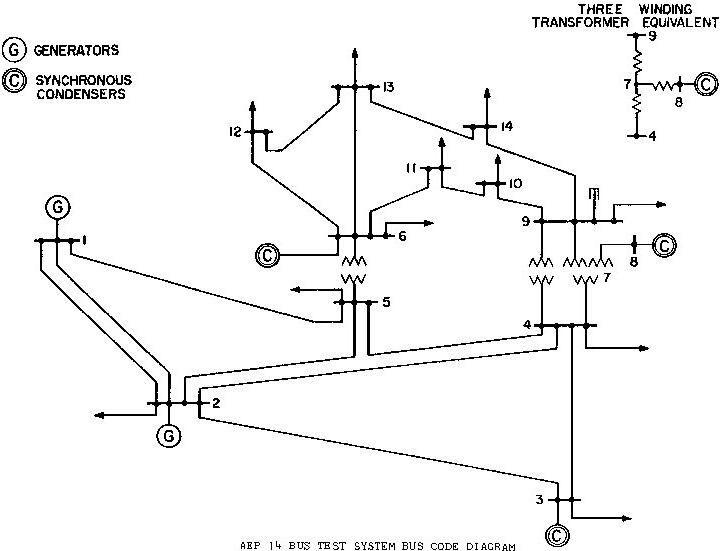}
		\caption{IEEE 14 Bus Single Line Diagram}
		\label{fig:IEEE14}
	\end{figure}
	
	In this section, we first describe an electric power grid scenario and highlight how it can be modeled as a graph. Then, we describe the MDCS problem, showcasing how solutions to it can help with sensor placement, for the unique monitoring of HVTs. Finally, we provide a quick overview of Moving Target Defense (MTD) and the notion of differential immunity.
	

	\subsection{The Electric Power Grid as a Graph}
	
	In \autoref{fig:IEEE14}, we show the IEEE 14 Bus single line diagram of an electrical power grid.
	In \cite{basu2018health}, the authors proposed a set of graph construction rules that model the monitoring of HVTs as a bipartite graph $G=(T \cup S, E)$, where $T$ represents the set of High Voltage Transformers (HVTs) that need to be uniquely monitored and $S$ represents the locations where the PMUs (or sensors) can be potentially placed (PMU's cannot be directly placed on HVTs), and $E$ represents the set of edges that exist if the operational behavior signal of an HVT ($t \in T$) reaches a PMU ($s \in S$) within a pre-specified number of hops. As Signal-to-Noise ratio (SNR) is used to measure the operational signal of an HVT in the real-world, and are known to quickly deteriorate over multiple hops, we, similar to prior works \cite{basu2018health,padhee2020identifying}, consider the number of hops to be at most $2$ (see \autoref{fig:14bus-graph}).
	

	\subsection{Minimum Discriminating Code Set (MDCS)}

	The MDCS problem is a special case of the Minimum Identifying Code Set (MICS) \cite{karpovsky1998new}, and was first studied in \cite{charbit2006discriminating}. Given a graph, the goal of MICS is to identify the smallest set of nodes on which sensors can be placed such that two properties are met (given domain-specific information propagation constraints). First, if an event occurs at an entity represented by a node in the graph, a unique set of sensors is activated leading to easy identification of the node (entity). Second, every node should trigger a non-empty set of sensors if an event occurs at the node. In MDCS, the problem is adapted to a bipartite graph scenario with two (disjoint) sets of nodes-- (i) nodes of interest, where an event may occur, which have to be uniquely identified with the sensors, and (ii) nodes on which sensors can be placed. Formally, we can define the MDCS problem in the context of sensor placement in power grid systems as follows \cite{basu2018health}.
	
	\begin{definition}
		Given a Bipartite Graph, $G = (T \cup S, E)$, a vertex set $S' \subseteq S$ is defined to be the Discriminating Code Set of $G$, if $\forall t \in T, N(t) \cap S'$ is unique, where $N(t)$ denotes the neighborhood of $t$. The {\em Minimum Discriminating Code Set} (MDCS) problem is to find the Discriminating Code Set of minimum size.
	\end{definition}
	
	\begin{figure}[t]
		\centering
		\includegraphics[width=\textwidth]{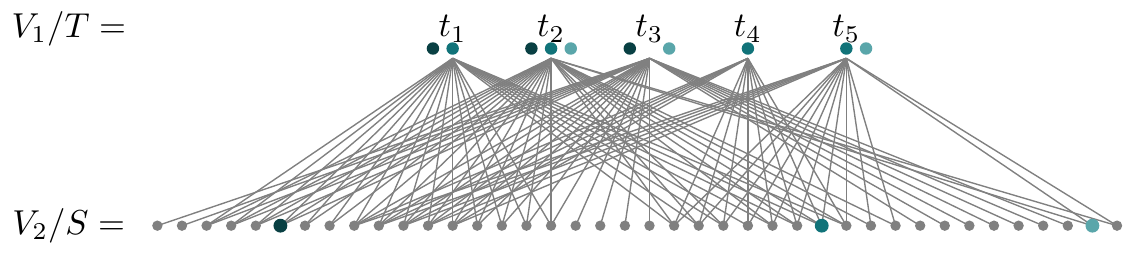}
		\caption{Bipartite Graph derived from the IEEE $14$-bus network with $2$-hop signal propagation constraints.}
		\label{fig:14bus-graph}
		\vspace{-12.0pt}
	\end{figure}
	
	\autoref{fig:14bus-graph} represents the bipartite graph obtained from \autoref{fig:IEEE14}, with $5$ nodes in $T$, representing the 5 HVTs, and $40$ nodes in $S$. An MDCS solution $S' \subseteq S$ of this graph consists of three nodes (indicated by the three colored nodes) which ensure that they provide a unique code to identify each of the $5$ nodes in $T$ (colors above the nodes of $T$ indicate the unique combination of sensors activated).
	

	\subsection{Moving Target Defense (MTD) and Differential Immunity}
	
	Conceptually, MTD, popular in cyber-security, seeks to continuously move between a set of system configurations available to a defender, to take away the attacker's advantage of reconnaissance \cite{jajodia2011moving}. The key idea is that the attacker may not encounter the expected system configuration at the time of the attack, thereby being rendered ineffective. Formally, an MTD system can be described using the three-tuple $\langle C, T, M \rangle$ where $C$ represents the set of system configurations a defender can move between, $T$ represents a timing function that describes when the defender moves and $M$ represents the movement strategy \cite{sengupta2020survey}.
	The goal of this work is two-fold-- (1) to construct a desirable set $C$ (for which we define the $K$-$\delta$MDCS problem in \autoref{sec:kdimcds}) and (2) an optimal movement strategy $M$ (by modeling the interaction as a game in \autoref{sec:mtdgame}).
	
	Note that when a single attack can cripple all the defense configurations $\in C$, MTD cannot aid in improving the robustness. In \cite{sengupta2019mtdeep}, the authors introduce the notion of {\em differential immunity} that aims at measuring the amount of diversity between configurations $\in C$. In this work, we consider a $C$ that is differentially immune (denoted as $\delta$), i.e. each attack, allowed by the threat model defined later, can only cripple one defense configuration. This ensures maximum diversity of $C$ and implies the highest robustness gains for the formulated MTD.
	
	
	
	\section{$K$ Differentially Immune MDCS ($K$-$\delta$MDCS)}
	\label{sec:kdimcds}
	
	To design the configuration set $C$ for an MTD system, we first need to find multiple MDCS sets of a bipartite graph. For this purpose, we desire $K$ differentially immune MDCS ($K$-$\delta$MDCS) where no two MDCS solutions share a common sensor placement point. Formally,
	
	\begin{definition}
		\textbf{($K$-$\delta$MDCS)} Given a Bipartite Graph, $G = (T \cup S, E)$, $K$ vertex sets $S_i \subseteq S, i \in \{1,\dots,K\}$ are defined to be $K$-$\delta$MDCS of $G$, if the following conditions hold-- (1) all the sets $S_i$ are MDCSs of graph $G$ and (2) for all possible pairs of sets $(S_i, S_j)$, $S_i \cap S_j = \emptyset$.
	\end{definition}
	
	First, we want to activate the minimum number of sensors placed in the network at any point in time. Hence, we use $K$ sets, all of which are MDCS, i.e. have the smallest cardinality. Second, the use of differentially immune MDCS tries to optimize for robustness in adversarial settings. If an attacker were to attack a particular sensor placement point $s \in S$, it can hope to, at best, cripple a singular MDCS $S_i\in C$, from uniquely identifying HVT failure. If the defender selects an MDCS $S_j \in C (j \neq i)$, then the attacker will not succeed in affecting the functionality of the power grid sensors. We will now show that the decision problem corresponding to $K$-$\delta$MDCS is NP-complete.
	
	\begin{figure}[t]
		\centering
		\includegraphics[width=\textwidth]{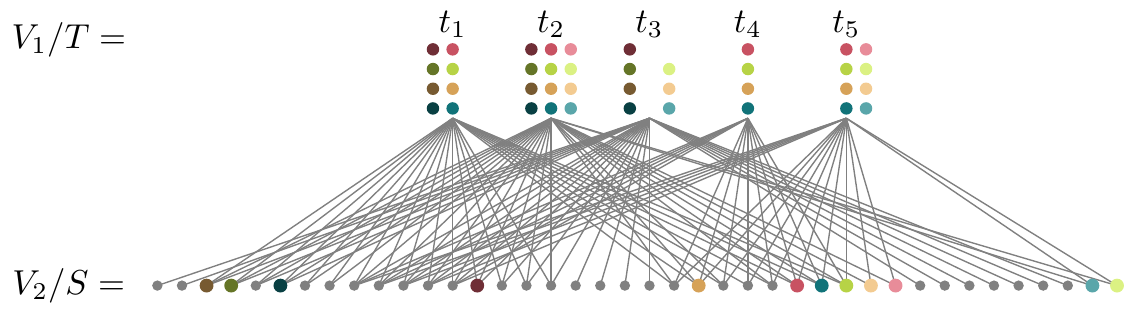}
		\caption{The IEEE 14-bus power grid graph has $4-\delta$MCDS solutions.}
		\label{fig:solutions}
	\end{figure}
	
	\begin{lemma}
		$K$-$\delta$MDCS is NP-Complete, given $K$ is an integer and $K>0$.
	\end{lemma}
	
	\paragraph{Proof.} We note that the original MDCS problem, which is known to be NP-Complete \cite{charbit2006discriminating}, is a special case (when $K=1$). \qed
	
	\begin{corollary}
		$K$-$\delta$ Graph Problems such as $K$-$\delta$Minimum Identifying Code Set (MICS), $K$-$\delta$Minimum Set Cover (MSC), $K$-$\delta$Minimum Vertex Cover (MVC) are NP-Complete when $K$ is an integer and $K > 0$.\footnote{Note that in the context of these problems, the distinction between the node sets $T$ and $S$ in MDCS are unnecessary and one can view the graphs as $G=(V,E)$.}
	\end{corollary}
	
	Let us denote the size of an MDCS for a bipartite graph $G$ as $m$. In $K$-$\delta$MDCS, the goal of the defender is to find $K$ MDCSs each of size $m$. Then, the defender needs to place $K*m$ sensors in the power grid and, at any point in time, activate an MDCS set (of size $m$) to uniquely identify failures in $T$. While a large number of defender strategies (i.e. larger values of $K$) helps to increase their options for sensor activation in turn reducing the success rate for the attacker, it also incurs the cost of placing $K*m$ sensors. Thus, the ideal choice of $K$ should trade-off robustness {\em vs.} sensor costs (when $K=1$, robustness using MTD is impossible to achieve).
	
	In cases where the defender has sufficient resources, one might ask {\em what is the maximum size of $K$?} Depending on the structure of the underlying graph, this question may have a trivial answer. For example, if the bipartite graph has a $t \in T$ and $N(t) = \{s\}, s \in S$, any MDCS of $G$ needs to place a sensor on $s$ to uniquely detect a fault in $t$. Hence, there can exist no two MDCSs that do not share a common node since $s$ has to be a part of both. In such cases, the max value of $K$, denoted as $K_{\max}$, is $1$. Beyond such cases, similar to the problem of finding the maximum value of $K$ in the $K$-clique problem, finding $K_{\max}$ demands a search procedure over the search space of $K$ that we now describe.
	
	\subsection{Finding max $K$ for $K$-$\delta$MDCS}
	
	We first propose a Quadratically Constrained Integer Linear Program (QCILP) that given a value of $K$, finds $K$ Discriminating Code Sets (DCSs). We then showcase the algorithm for searching over possible values of $k \in \{1,\dots,|S|\}$ to find the largest $K$. To define the QCILP for $G = (T \cup S, E)$, we first consider $|S|*k$ binary variables where, $x_{sk} = 1$ if a sensor is placed in node $s \in S$ for the $k$th DCS, and $0$ otherwise. We also use a variable $l$ that denotes the size of the DCSs found. We can now describe our QCILP, presented below.
	
	\begin{eqnarray}
	\min_{l,x} && l \label{eqn:opt} \\
	s.t. \quad l &=& \sum_{s} x_{sk} \quad \forall k \quad \textsf{\scriptsize \color{RoyalBlue!50!white} All $k$ DCS has the same size $l$.}\nonumber \\
	\sum_{s \in S} (x_{sk} - x_{sk'})^2 &=& 2l \quad \forall (k,k') \quad \textsf{\scriptsize \color{RoyalBlue!50!white} No two DCSs should have a common sensor.} \nonumber \\
	\sum_{s \in N(t)} x_{sk} &\geq& 1 \quad \forall t, \forall k \qquad \textsf{\scriptsize \color{RoyalBlue!50!white} All $t \in T$ has a sensor monitoring them for all the $k$ solutions.}\nonumber \\
	\sum_{s \in N(t) \Delta N(t')} x_{sk} &\geq& 1 \quad \forall (t,t'), \forall k \quad \textsf{\scriptsize \color{RoyalBlue!50!white} $t$ and $t'$ trigger unique sensors for the $k$-th DCS.} \nonumber \\
	x_{sk} &\in& {0, 1} \forall s, \forall k \nonumber
	\end{eqnarray}
	
	The last two constraints ensure that each of the $K$ solutions is Discrimination Code Sets where (1) all $t \in T$ trigger at least one sensor $s \in S$ and (2) for all pairs of $t$ and $t'$ (both $\in T$), there exists at least one sensor in the symmetric difference set of $t$ and $t'$ that is a part of the DCS, which in turn uniquely distinguishes between $t$ and $t'$. The first two constraints ensure that all $k$ DCSs are of equal size and no two DCSs shares a common sensor. We can now ask the question as to whether the DCSs found by \autoref{eqn:opt} is indeed the Minimum DCSs (MDCSs) for the graph $G$. In this regard, we now show the following.
	
	
	\begin{theorem}
		For all values $K \leq K_{\max}$, the optimization problem in \autoref{eqn:opt} returns $K$-$\delta$MDCS.
	\end{theorem}
	
	\paragraph{Proof.} We consider proof by contradiction. Given the value of $K (\leq K_{\max})$, let us assume that the solution returned by \autoref{eqn:opt} is not the $K$-$\delta$MDCS for the graph $G$. If this is the case, at least one of the two properties in the definition of $K$-$\delta$MDCS is violated. Thus, either (1) the returned solution consists of a DCS that is not the Minimum DCS, or (2) there exists a sub-set (of size greater than one) among the set of DCSs that share a common node.
	
	Owing to the third and fourth constraints, all the solutions constitute a DCS. Now, if (1) is violated, all the DCSs returned by the QCILP, of length $l$, are not the MDCS for $G$. Thus, the MDCS must have a DCS of size $l' \leq l$. Given that the minimization objective finds the smallest DCS and $K \leq K_{\max}$, this cannot be possible. Hence, (1) does not hold.
	
	For (2), let us say that there exists a subset of the DCSs returned that share a common node. If this was the case, then at least one solution pair has to share a common node. If this node is denoted as $s^*$ and the two solutions are termed as $k$ and $k'$, then for the second constraint, given $x_{s^*k} = x_{s^*k'} = 1$, the term for $s^*$ is zero. Even if the other $l-1$ nodes in the solutions $k$ and $k'$ are unique, the terms will add up to $2 * (l-1)$ thereby violating the second constraint. This is not possible and as a consequence, (2) does not hold. \hfill \qed
	
	Given this, we can now consider cases where $K > K_{\max}$. When $K > K_{\max}$, the optimization problem in \autoref{eqn:opt} is either infeasible or returns $K$ DCSs that are not MDCS for graph G. This condition holds by the definition of $K_{\max}$ (proof by contradiction ensues if neither of the two cases holds). With these conditions in mind we can design an iterative approach, shown in \autoref{algo:sql}, to find the $K_{\max}-\delta$MDCS of a given graph.
	
	\begin{algorithm}[t]
		\scriptsize
		\begin{algorithmic}[1]
			\STATE{\textit{In:} $G = (T \cup S, E)$}
			\STATE{\textit{Out:} $K_{\max}-\delta$MDCS}
			\STATE{solutions $\leftarrow \emptyset$}
			\STATE{$K \leftarrow 1$}
			\WHILE{$K \leq |S|$}
			\STATE{solutions$_K \leftarrow$ Solve \autoref{eqn:opt} with $K$}
			\IF{solutions$_K == \emptyset$}
			\STATE{break \qquad\qquad\qquad\qquad\qquad\qquad\qquad \textsf{\color{RoyalBlue!50!white} Infeasible for $K > K_{\max}$}}
			\ENDIF
			\IF{solutions $ != \emptyset$ \AND $|$solutions$(l)| < |$solutions$_K(l)|$ }
			\STATE{break \qquad\qquad\qquad\qquad\qquad\qquad\qquad \textsf{\color{RoyalBlue!50!white} DCS returned is not MDCS for $K > K_{\max}$}}
			\ENDIF
			\STATE{solutions $\leftarrow$ solutions$_K$}
			\STATE{$K \leftarrow K+1$}
			\ENDWHILE
			\RETURN solutions
		\end{algorithmic}
		\caption{Finding $K_{\max}-\delta$MDCS.}
		\label{algo:sql}
	\end{algorithm}
	
	\autoref{fig:solutions} showcases the $4-\delta$MDCS solutions returned by \autoref{algo:sql} for the 14-bus power grid network. The different colors indicate the different MDCSs found for $G$ and the shades of the same color indicate an MDCS set. As shown, each of the four MDCS has a size of $l=3$ and uniquely identifies all the transformers $T$. The lack of overlapping colors in the bottom set of nodes indicates that no two MDCS share a common $s \in S$.
	
	While the procedure in \autoref{algo:sql} finds the $K_{\max}-\delta$MDCS, it can be time-consuming for the largest networks (although it works well on large power-grids as shown in the experimental section). Thus, one can consider a greedy approach in which one solves the MDCS problem using \cite{basu2018health}. We then solve this ILP with the additional constraints that $x_s = 0$ for all the sensors found in the current solution and keep doing so until (1) the ILP becomes infeasible or (2) results in DCS that does not have minimum cardinality. In the experimental section, we will see that although this approach is faster, it can output $K$-$\delta$MDCS where $K < K_{\max}$. The sub-optimality is a result of the ordering ``enforced'' by the current optimal MDCSs which in turn, proves to be infeasible constraints for the latter iterations of the problem.

	\section{Game Theoretic Formulation}
	\label{sec:mtdgame}
	
\begin{figure}[t]
		\centering
		\includegraphics[width=\textwidth]{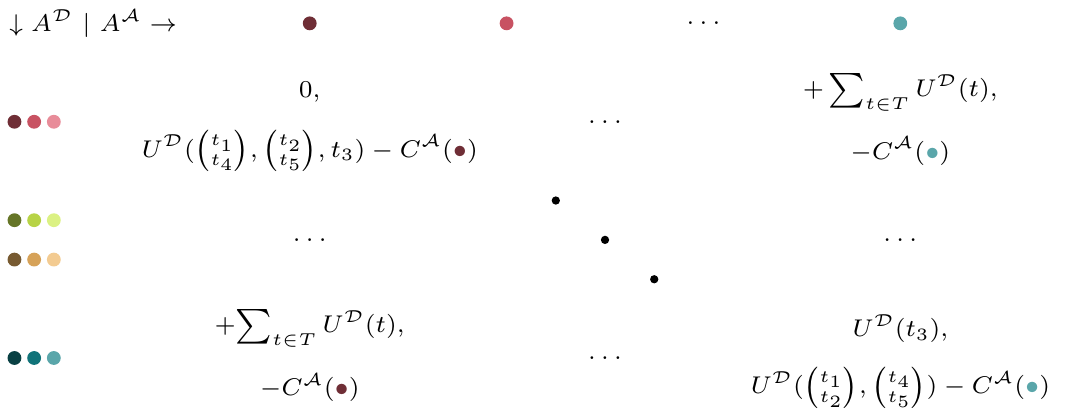}
		\caption{Game-matrix for the dynamic sensor activation problem.}
		\label{fig:game}
		
	\end{figure}

	The defender's goal is to maintain the unique identifying capability of HVTs at all times. Conversely, the attacker tries to prevent this capability, thereby making it harder for the defender to effectively monitor the HVTs. Here, we seek to find the optimal movement function $M$ for the sensor activation MTD to aid the defender to realize its objective. To do so, we consider a strong threat-model where the attacker $\mathcal{A}$ with recon, is aware of the defender $\mathcal{D}$'s (probabilistic) sensor activation strategy, thereby making the Stackelberg Equilibrium an appropriate solution concept for our setting. We use a polynomial-time approach to find the Strong Stackelberg Equilibrium of the game \cite{conitzer2006computing}.
	We now briefly describe the various parameters of the formulated game (see \autoref{fig:game}).
	
	\paragraph{Defense Actions} The defender has $K_{\max}$ pure strategies and the configuration set $C = K_{\max}-\delta$MDCS. If one uses the greedy algorithm instead of the optimal approach (both described in the previous section), the number of pure strategies obtained may be less than $K_{\max}$.
	
	\paragraph{Attack Actions} We assume that an attacker can spend reconnaissance effort in figuring out the sensor placement point. Thus, its action set includes attacking a sensor that may be considered for activation (instead of all nodes in $|S|$). While one can consider attackers with the capability to attack multiple sensor activation points, it is often too expensive a cost model as it demands resource procurement and distribution over a wide geographic area.
	
	\paragraph{Player Utilities} The game has two different kinds of utilities that are used to calculate the rewards. First, the defender receives the utility associated with uniquely identifying a transformer $t \in T$ in the case of anomalous spikes indicative of failure (to occur). We assume that a transformer supplying power to an important building (eg. the White House or the Pentagon) is considered to be more important than one supplying power to a residential area. Second, the attacker's cost for attacking a particular sensor needs to be considered. While some sensors may be placed in high-security areas, others may be easier to access. We conduct randomized trials with both these values $\in [0,10]$, with $10$ indicating the HVT/sensor most important to protect/difficult to attack.
	
	In the bottom right corner of \autoref{fig:game}, the defender, owing to the attacker attacking a sensor, is only able to uniquely identify $t_3$ and thus, only gets reward proportional to it. Contrarily, the attacker, due to attacking a sensor, can make failures of $t_1$ and $t_2$ (and $t_4$ and $t_5$) indistinguishable and receives the corresponding utilities, minus the cost of attacking the sensor denoted by the light blue node ($\in S$, \autoref{fig:solutions}). Similarly, if the attacker selects the attack represented by the first attack column (sensor denoted by the dark brown node), the defender cannot identify any HVT and thus, gets a utility of zero.
	
	\begin{table}[t]
		\scriptsize
		\centering
		\caption{Game parameters and defender's reward for playing the different $C$s and $M$s for the various power-grid networks.}
		\label{tab:table1}
		\begin{tabular}{@{}l>{\centering\arraybackslash}p{1.25cm}>{\centering\arraybackslash}p{1.25cm}>{\centering\arraybackslash}p{1.3cm}>{\centering\arraybackslash}p{1.42cm}>{\centering\arraybackslash}p{1.42cm}>{\centering\arraybackslash}p{1.42cm}>{\centering\arraybackslash}p{1.42cm}}
			\toprule
			& &  $C$ & & \multicolumn{4}{c}{Movement Function $M$}\\
			\cmidrule{5-8}
			Graph & $|S|+|T|$ & $|A^\mathcal{D}|$ $(K/K_{\max})$ & $|A^\mathcal{A}|$  $(K/K_{\max})$ & URS $~~~(K)~~$ & URS $(K_{\max})$ & SSE $~~~(K)~~$ & SSE $(K_{\max})$\\
			\midrule
			14 Bus & 45 & 4/4 & 12/12 &  18.5$\pm$4.7& 18.65$\pm$4.7 & 20.62$\pm$4.6 & \textbf{20.72}$\pm$4.6\\
			30 Bus & 89 & 4/4 & 16/16 & 26.45$\pm$5.7 & 27.25$\pm$5.6 & 29.44$\pm$6 &\textbf{29.9}$\pm$5.8\\ 
			39 Bus & 96 & 7/9 & 28/36 & 18.7$\pm$5 & 19.24$\pm$5.2 & \textbf{19.8}$\pm$5.3&19.73$\pm$5.3\\
			57 Bus & 170 & 6/6 & 60/60 & 70.76$\pm$10.8 & 70.88$\pm$11.1 & \textbf{73.5}$\pm$10.6& 73.07$\pm$10.7\\
			89 Bus & 422 & 16/21 & 96/126 & 50.67$\pm$8.9 &51$\pm$9  & \textbf{52.2}$\pm$9.2 & \textbf{52.2}$\pm$9.2\\
			118 Bus &367 & 2/2 & 10/10 & 31.35 $\pm$6 & 31.6 $\pm$ 6 & 32.45$\pm$6.4 & \textbf{32.61}$\pm$6.1\\
			2383 Bus & 5927 & 2/3 & 212/318 & 832.7$\pm$38.7 & 836.16$\pm$36.7 & 835.34$\pm$39 & \textbf{842.34}$\pm$39.4\\
			\bottomrule
		\end{tabular}
		
	\end{table}

	\section{Experimental Simulation}

	In this section, we conduct simulation studies on seven IEEE test graphs popular in the power domain \cite{zimmerman2010matpower}. Characteristics of these graphs such as the total number of nodes (i.e. $|S|+|T|$) are shown in \autoref{tab:table1}. The table further lists the $K$ values for the $K$-$\delta$MDCS found by the greedy and the optimal \autoref{algo:sql}, and is denoted by $K$ and $K_{max}$ respectively. The number of attacker strategies is listed in the fourth column. This value can be obtained by multiplying the corresponding $K$ value with the size of an MDCS for graph $G$, since none of the $K$-$\delta$MDCS share a common node. We now discuss two results-- (1) the effectiveness of the game-theoretic equilibrium compared to the Uniform Random Strategy baseline (which chooses to activate a particular MDCS with equal probability) and (2) the time is taken by the greedy and the optimal algorithm and their respective solution quality. \footnote{The code for the experiments can be found at \url{https://github.com/kaustav-basu/Robust-MICS}}

	\subsubsection*{Effectiveness of Game-Theoretic Equilibrium}
	In \autoref{tab:table1}, we show that in all test cases, the optimal movement strategy at the Strong Stackelberg Equilibrium (SSE) gives the defender a higher reward than choosing URS. When using URS or SSE, in most cases we see higher gains when the construction of the MTD configuration set $C$ is optimal ($URS(K_{\max})$ obtained from \autoref{algo:sql}) as opposed to using a greedy algorithm ($URS(K)$). We expected this as the higher number of differentially immune options (as $K_{\max} > K$) chosen with equal probability reduces the probability of picking the weakest strategy. When the value of $K_{\max} = K$, such as for 14, 30, 57 and 118 buses, we see that the difference between the two versions of URS (or two versions of SSE) are negligible. A reason for the non-zero difference between the rewards values arises because of the MDCS sets chosen, although the total number of sets chosen are the same. We also see that the difference in defender rewards can be large even when the difference between $K$ and $K_{\max}$ is small in the case of larger networks (eg. 2383 bus). Thus, without finding the $K_{\max}$ and the SSE for the optimal $C$, it is hard to establish the loss in rewards. Given that these strategies are pre-computed, the power grid utility operator should not consider the greedy strategy unless the time required becomes prohibitive.
	
	\begin{figure}[t]
		\centering
		
		\includegraphics[width = 0.9\textwidth]{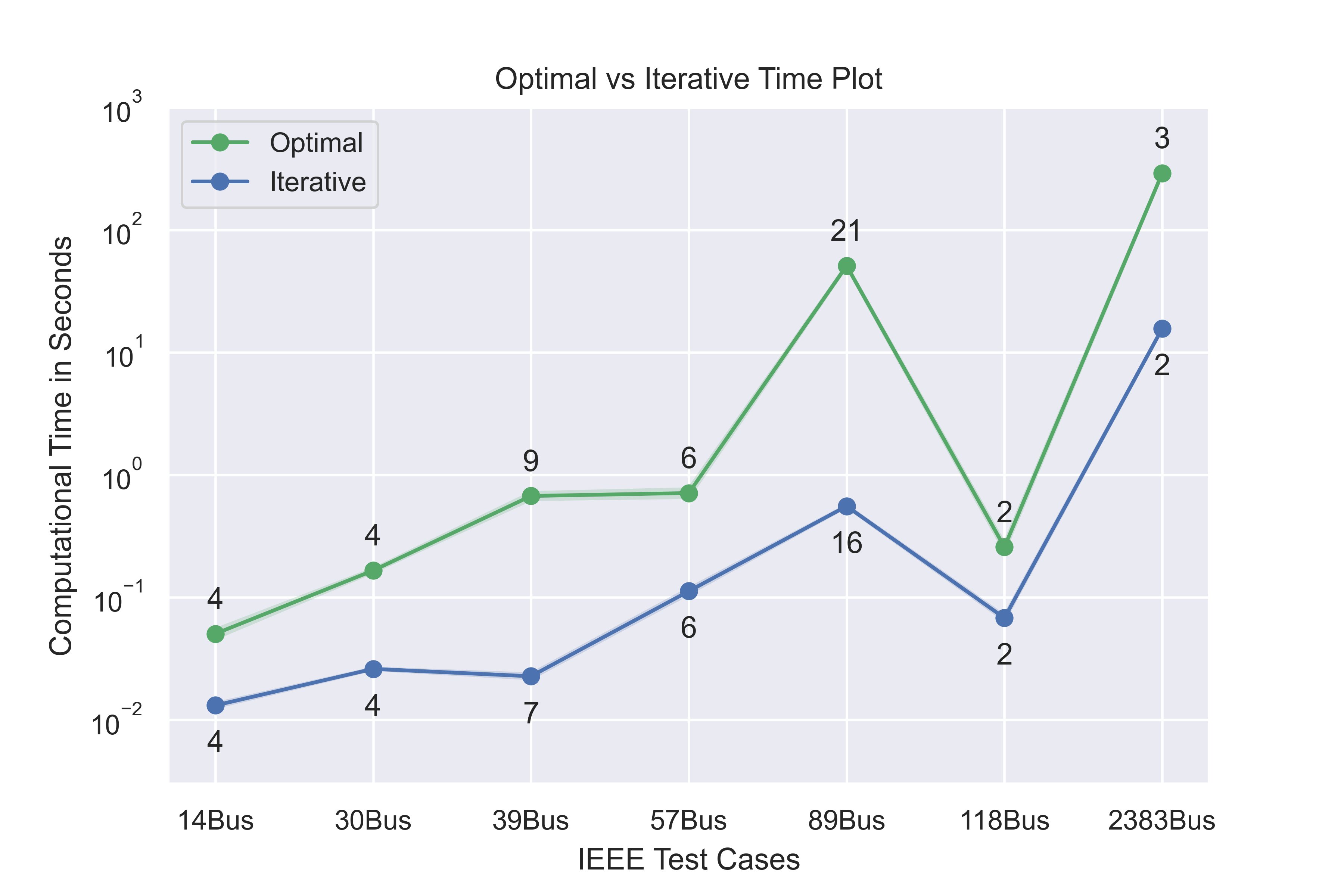}
		\caption{Time taken by the optimal (\autoref{algo:sql}) {\em vs.} the greedy approach for finding $K_{\max}-\delta$MDCS and $K$-$\delta$MDCS (the $K$ values are shown above the plot points).}
		\label{fig:plot}
		
	\end{figure}

	\subsubsection{Computational Time for finding $C$}
	
	In \autoref{fig:plot}, we compare the time taken for finding the configuration set $C$ using the optimal $vs.$ the greedy approach. We choose the logarithmic scale for the y-axis because the computational time of the optimal and greedy approaches for the 14, 30, 39, 57, and 118 buses was less than a second, and thus difficult to distinguish between on a linear scale.
	The largest disparity occurs when the size of the optimal set $K_{\max}$ is greater than the $K$-sized set found by the greedy approach (39/89/2383 Bus). In other cases, while the optimal approach is slower, it provides the guarantee that no solution with a greater $K$ exists, which is absent in the greedy case. A case where the logarithmic scale, from a visualization perspective, does not do justice is the 2383-Bus. The time taken by the greedy approach is $15s$ compared to $291s$ taken by the optimal approach. While the $K$ value differs by a factor of one, the resultant gain in defender's game value, as shown in \autoref{tab:table1}, is relatively large. Thus, the added time in generating the optimal configuration set needs to be criticized based on the gain obtained in the underlying game.
	
	We also consider the pragmatic scenario when the $K$ value is fixed by the defender up-front owing to budget restrictions of sensors that can be placed in the power network. In this case, the greedy approach has to iteratively find one solution at a time, adding them to the constraint set of future iterations until the desired $k$ is reached. On the other hand, the iterative procedure in \autoref{algo:sql} can be altogether ignored and one can simply return the solution found by the optimization problem in \autoref{eqn:opt}.

	\section{Related Works}

	Adversarial attacks on power grids comprise of false-data injection, jamming, DoS and packet-dropping attacks \cite{deka2015optimal,deng2016false,n2020modelagnostic}. While researchers have proposed a multitude of defense mechanisms \cite{tan2017survey}, including Moving Target Defense (MTDs) \cite{chatfield2017moving,potteiger2020security}, they do not consider the problem of sensor placement to monitor HVTs. On the other hand, works that leverage the formalism of Discriminating Code Sets \cite{charbit2006discriminating} to optimize sensor placement \cite{basu2018health}, have focused on scalability issues and provided theoretical bounds in these settings \cite{basu2019sensor}; completely ignore the issue of robustness to adversarial intent. In this work, we attempted to fill in this gap.
	
	
	
	While an array of research work has formally investigated the notion of finding an optimal movement function $M$ for MTDs, the configuration set $C$ is pre-decided based on heuristic guidance from security experts \cite{sengupta2020survey}. While some works consider the aspect of differential immunity by analyzing code overlap for cyber systems \cite{carter2014game} or Jacobians of gradients for deep neural networks \cite{adam2018stochastic}, these measures have no way of ensuring differential immunity. The notion of k-set diverse solutions in Constraint Satisfaction Programming (CSP) \cite{hebrard2005finding}, although conceptually similar to our notion of differential immunity, does not have the added constraint of finding a minimum sized solution (as in the case of MDCS). In adversarial scenarios, our work is the first to formalize the notion of diversity in graphs and propose linear programming methods to find them.
	
	

	\section{Conclusion}

	We considered the problem of monitoring the behavior of HVTs in adversarial settings and proposed an approach based on MTD, formulating it as a game between the power utility company (the defender) and an adversary. We showed that finding the configuration set for the defender is NP-Complete and presented two algorithms-- an optimal QC-ILP and a greedy iterative-ILP. Optimal movement strategies at Stackelberg Equilibrium enabled the defender to activate $k$ sensors at a time and uniquely identify failure points in the face of adversarial attacks. Results obtained on several IEEE test cases showed that the proposed methods yields the highest expected reward for the defender.
	
	\subsubsection*{Acknowledgements} The research is supported in part by ONR grants N00014-16-1-2892, N00014-18-1-2442, N00014-18-1-2840, N00014-19-1-2119, AFOSR grant FA9550-18-1-0067, DARPA SAIL-ON grant W911NF-19-2-0006, 
	and DARPA CHASE under Grant W912CG19-C-0003 (via IBM).
	
	\bibliographystyle{splncs04}
	\bibliography{ref}
	
\end{document}